\documentclass[showpacs,preprintnumbers,amsmath,amssymb,floatfix,prl,superscriptaddress]{article}
\usepackage[latin1]{inputenc} 
\usepackage{setspace}
\usepackage{graphicx}
\usepackage{graphics}
\usepackage{subfigure}
\usepackage{dcolumn}
\usepackage{amsmath}
\usepackage{geometry}
\usepackage{amssymb}
\usepackage{bm}
\usepackage{color}
\usepackage{SIunits}
\usepackage[normalem]{ulem}

\begin{document}
\title{Optical Properties of Relativistic Plasma Mirrors}

\author{H. Vincenti$^{1}$,  S. Monchoc\'e$^1$, S. Kahaly$^1$, Ph. Martin$^1$ and F. Qu\'er\'e$^{1,\dag}$}
\date{\today}
\maketitle

\small{
\noindent $^1$Service des Photons, Atomes et Mol\'{e}cules, CEA, DSM/IRAMIS, CEN Saclay, 91191 Gif-sur-Yvette, France
}
\\
\let\thefootnote\relax\footnotetext{$^\dag$Corresponding author:fabien.quere@cea.fr}

\begin{abstract}
The advent of ultrahigh-power femtosecond lasers creates a need for optical components suitable to handle ultrahigh light intensities. Due to the unavoidable laser-induced ionization of matter, these components will have to be based on a plasma medium.  An archetype of such optical elements is a plasma mirror, created when an intense femtosecond laser pulse impinges on a solid target. It consists of a dense plasma, formed by the laser field itself, which specularly reflects the main part of the pulse. Plasma mirrors have major potential applications as active optical elements to manipulate the temporal and spatial properties of intense laser beams, in particular for the generation of intense attosecond pulses of light. We investigate the basic physics involved in the deformation of a plasma mirror resulting from the light pressure exerted by the ultraintense laser during reflection, by deriving a simple model of this fundamental process, which we validate both numerically and experimentally. The understanding of this deformation is essential for all future applications of plasma mirrors, especially for the generation of collimated attosecond beams. We show how its effect on the attosecond beam divergence can be mitigated by using the laser phase, thus providing crucial control for future applications in attosecond science.
\end{abstract}

\maketitle

\par
Ultrafast laser technology now makes it possible to study the interaction of femtosecond (fs) laser pulses with plasmas in an extreme regime, where the motion of electrons in the laser field is relativistic~\cite{MourouReview}. With several facilities aiming at peak powers beyond a PetaWatt, the study of new regimes of quantum electrodynamics should thus become feasible in the near future ~\cite{DiPiazza}. The rapid growth in the number of high-power ultrashort lasers is also driven by the perspective of societal and scientific applications, such as compact laser-driven particle accelerators~\cite{DaidoRPPion,Malka,Fuchs}.

\par
These laser developments and their prospects call for new types of optical elements, that can be used to manipulate and tailor ultrahigh-power laser beams at very high intensities $I$, both in the temporal and spatial domains. As soon as $I \gtrsim 10^{13}$ $W/cm^2$, any medium gets strongly ionized by the field, making conventional optics inappropriate: in this regime, optical components will inevitably consist of a plasma medium. 
Easy to use and versatile, plasma mirrors (PM) have a major role to play as high-intensity optical components~\cite{article_NP}, and constitute simple testbeds for models of relativistic laser-plasma interaction. 

PM are already routinely used at moderate light intensities ($10^{14}-10^{16}$ $W/cm^2$) as ultrafast optical switches, to enhance the temporal contrast of femtosecond lasers~\cite{doumy, dromey_mp, monot, article_NP}.
As $I \gtrsim 10^{16}$ $W/cm^2$, the non-linear response of PMs to the laser field results in sub-cycle temporal modulations of the reflected field, associated to high-order harmonic generation (HHG) in its spectrum~\cite{teubner_RMP, FabienJPB2010}. These harmonics, generated through different mechanisms, are associated in the time domain to attosecond pulses~\cite{plaja_roso,nomura_as}. Beyond $\sim 10^{18}$ $W/cm^2$, a key HHG mechanism is the Relativistic Oscillating Mirror (ROM), where the laser-driven oscillation of the plasma surface induces a periodic Doppler effect on the reflected field~\cite{Bulanov1994,Lichters1996,Baeva2006,FabienJPB2010,Gonoskov2011}, which can result in harmonic orders of several thousands~\cite{dromey_NP}. Plasma mirrors thus hold great promise for the generation of intense attosecond pulses of light~\cite{tsaki,Sansone}, which would break down a major barrier in attosecond science, opening the way to potential ground-breaking applications such as pump-probe experiments on electron dynamics in matter~\cite{KrauszRMP}.

\par
In addition to these temporal effects, the initial solid target on which the PM is created can be geometrically shaped, to also spatially manipulate the reflected beam. At moderate intensities, elliptical PMs have thus recently allowed extremely tight focusing of a high-power laser beam~\cite{Motoaki}. In the relativistic regime, curved PMs have been proposed as a way to focus the very high generated harmonic orders to a spot size $w <<\lambda_L$ (where $\lambda_L$ is the laser wavelength)~\cite{Gordienko_focusing, Gonoskov2011}. Combined with their attosecond temporal bunching, this is a promising path to boost the peak intensity of ultrashort lasers, which might help approaching the Schwinger limit~\cite{SchwingerPR} $I_s=2.25\cdot10^{29}$ $W/cm^2$, where the light field starts inducing electron-positron pair creation from vacuum~\cite{BellSchPRL,BulanovSchPRL}.

\par
In these high intensity applications, the laser field exerts such a high pressure on the plasma (typically $5$ Gbar for $I\approx10^{19}$ $W/cm^2$) that it induces a significant motion of the PM surface, even during a femtosecond laser pulse. Any spatial variation of the intensity on target, as generally occurs at or around focus, then leads to a deformation of the PM surface -typically a curvature- which can affect the spatial~\cite{dromey_diff, dromey_cwe, Horlein} and spectral~\cite{anDerBrugge} properties of the reflected beam. Beyond its fundamental interest, understanding and controlling this intrinsic dynamics of PM is crucial for any of the previous applications. It in particular determines the divergence of attosecond beams produced from plasma mirrors, which is a key parameter for future experiments.



In this article, we elucidate the physics of the light-induced curvature of the PM, with an analytical model of the surface dynamics and its consequences on the reflected light. Despite its simplicity, it captures the essential aspects of this process, and disentangles the influences of electron and ion dynamics in the femtosecond regime. Due to their small wavelengths high-order harmonics generated on PM are strongly affected, and thus constitute sensitive probes of its curvature. We present some of the most exhaustive measurements of the ROM harmonic properties performed to date, which we use to validate this model experimentally. Controlling the spatial properties of these harmonics is crucial for future applications in attosecond science. We finally demonstrate that such a control can be achieved very simply by using the spatial phase of the driving laser. 

\section*{Model of laser-induced plasma mirror curvature} 

Properly describing the PM surface motion requires taking into account both the plasma electron and ion dynamics. The response time of electrons to the laser field is much smaller than the  optical period, while ions react on a longer time scale due to their larger mass. Akin to the Born-Oppenheimer approximation in molecular physics, this makes it possible to model the system in three steps: (i) we first describe the quasi-instantaneous response of electrons to the laser field, considering a given ion background (Fig.\ref{sim1}(a)); (ii) we then calculate the slow ion motion, resulting from the combined actions of the laser-field and of the charge separation fields it induces (Fig.\ref{sim1}(b)) and (iii) finally the influence of the slow dynamics on the fast one is included, to determine the surface motion over the entire laser pulse (Fig.\ref{sim1}(c)). The derivations of all formulas and their validation by Particle-In-Cell (PIC) simulations are provided in the online supplementary information.

\subsection*{Electron dynamics}
\par

\begin{figure*}[t]
\centering \includegraphics[scale=0.85]{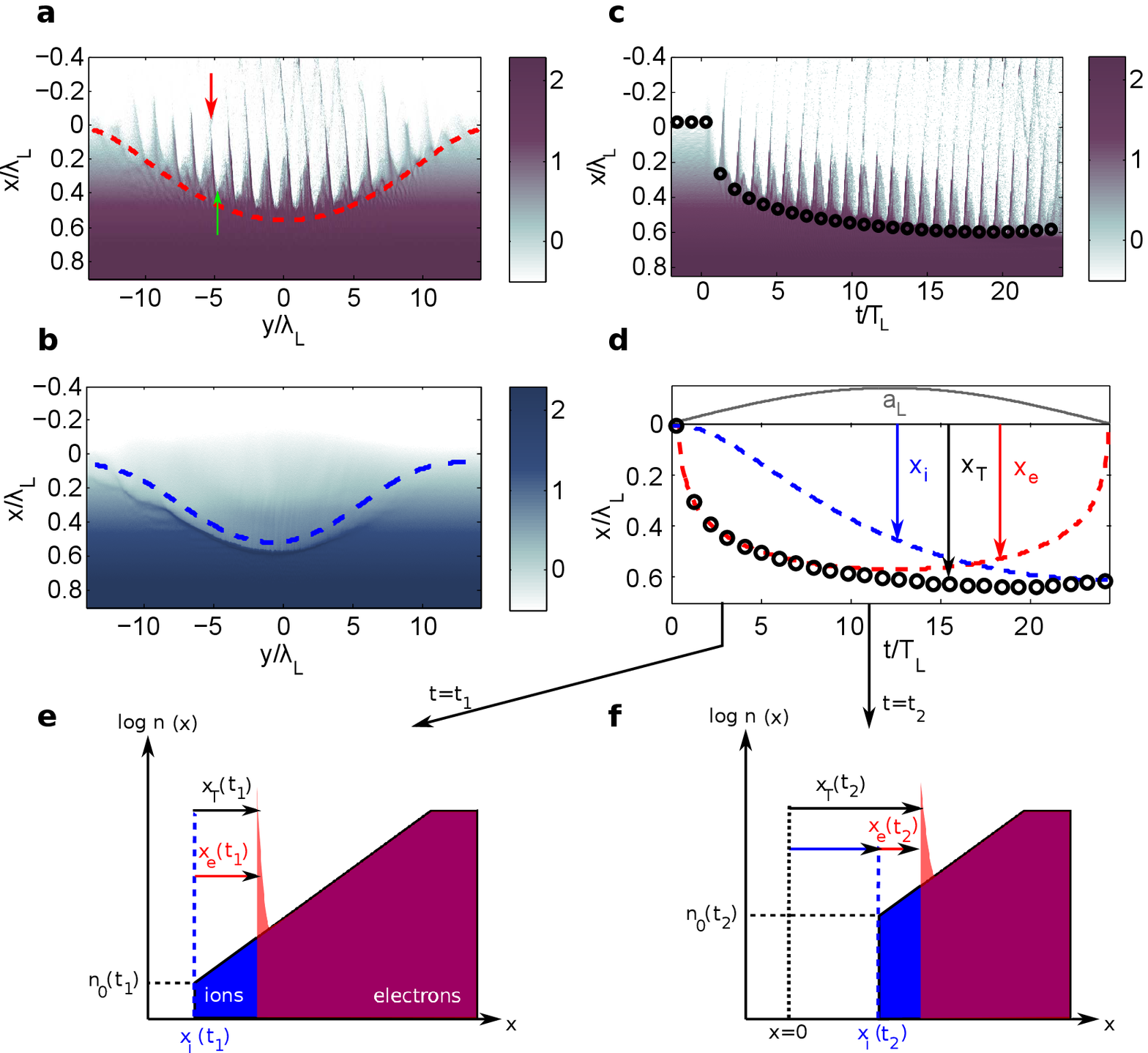}
\caption{\textbf{Laser-induced curvature of a relativistic plasma mirror.} (a) Spatial map of the plasma electron density $n_e(x,y)$ at the maximum of a laser pulse with a gaussian focus, from a 2D PIC simulation ($a_L=8, L=\lambda/8,\theta=45^o$) performed with fixed ions. (b) Same spatial map, now for the plasma ion density $n_i(x,y)$ (charge state of the ions $Z=1$) at the maximum of the laser pulse, from a 2D PIC simulation with moving ions in the same interaction conditions as in (a). (c) Temporal evolution of the electron density $n_e(x,t)$ at the center $y=0$ of the surface, with moving ions. A logarithmic scale is used in all cases, and densities are expressed in units of $n_c$. In (a) and (b), the dashed curves show the predictions of the model (Eq.(\ref{xe}) in (a), Eq.(\ref{xpL}) in (b)). The results of the total model for $x_T(t)=x_i(t)+x_e(t)$, that combines electron and ion dynamics, are shown by the black dots in (c). In (d), this model is used to disentangle the contributions of the electron dynamics and ion dynamics to the total surface displacement, by plotting: (i) the electron surface displacement $x_e^{FI}$ when no ion motion occurs, calculated using Eq.(\ref{xe}) with a fixed value $n_0=n_c cos^2 \theta$, (ii) the ion surface displacement $x_i$, calculated using Eq.(\ref{xpL}), and (iii) the total displacement $x_T=x_i+x_e$, where $x_e \neq x_e^{FI}$ is now the electron surface displacement calculated when ion motion is taken into account, by using $n_0=n(x_i)$ in Eq.(\ref{xe}). Panel (e) and (f) sketch the electron (red) and ion (blue) density profiles, at two different times of the laser pulse, and define the different quantities used in the model.  All surface displacements are calculated with respect to the reference position $x_e=x_i=0$, where the laser field reflects at the very beginning of the laser pulse, at low intensity. For an angle of incidence $\theta$, this is the point where $n=n_c \cos^2 \theta$, which also corresponds to the value of $n_0$ in Eq.(\ref{xe}) at the beginning of the interaction.
}  
\vskip -0.5cm
\label{sim1}
\end{figure*}

Qualitatively, the plasma electrons respond to the laser field as a spring, being alternatively pushed inside, and pulled outside of the ion background in each optical period~\cite{Gonoskov2011}. When pulled outward, they form relativistic electron jets (red arrow in Fig.\ref{sim1}(a)), that are responsible for the ROM attosecond pulse emission. When pushed inward, a high-density spike is formed at the sharp surface of the electron distribution (green arrow in Fig.\ref{sim1}(a)), at a position $x_e(t)$ (Fig.\ref{sim1}(e-f)). 
A detailed analysis of PIC simulations (see supplementary information) shows that the position of the outgoing electron jet responsible for the emission of an attosecond pulse in each laser cycle is tied to the position of the high-density spike formed in this compression phase, and thus follows the same evolution as the laser intensity changes in time or space. We therefore concentrate on the value of $x_e(t)$, which can be easily determined by the balance between the pushing force exerted by the laser field, and the restoring force exerted by the ion background. In the relativistic regime, this balance leads to the following expression for the maximum inward excursion $x_e$ of electrons in a given optical period:
\begin{equation}
x_e=L\ln\left[1+\frac{2\lambda_L a_L(1+\sin\theta)}{2\pi L }\frac{n_c}{n_0}\right]
\label{xe}
\end{equation}
where $\theta$ is the angle of incidence of the laser on the PM, and $n_c$ is the critical plasma density at the laser frequency. $n_0$ is the ion charge density at the ion-vacuum boundary (Fig.\ref{sim1}(e-f)), i.e. the density from which the laser field starts pushing electrons inside the ion background. For this derivation, the ion density gradient at the PM surface has been assumed to be exponential beyond $n_0$, with a scale length $L$, i.e. $n(x) \propto \exp(x/L)$ for $n>n_0$ (Fig.\ref{sim1}(e-f)). $L$ is a crucial parameter of the interaction, which in particular strongly affects the HHG efficiency~\cite{FabienJPB2010,Rodel_grad,KahalyGradient}. $x_e$ increases for larger values of $L$ in Eq.(\ref{xe}), because the laser field can more easily push electrons inside a smoother ion background. 

The electron boundary displacement $x_e$ also increases with $a_L=eA_L/m_ec=\left[I(W.cm^{-2})\lambda_L^2(\mu m^{2})/1.37\cdot10^{18}\right]^{1/2}$, the amplitude of the normalized vector potential of the incident laser field: the higher this amplitude, the further electrons get pushed inside the target. For a focused laser pulse, the field envelop is a function of both time and space, $a_L(y,t)$. 
The spatial envelop results in an overall spatial curvature -a denting- of the plasma electron density surface. This laser-cycle-averaged curvature is clearly observed on a spatial map of electron density at $t_0$ corresponding to the laser pulse maximum  (Fig.\ref{sim1}(a)). It is very well reproduced by the curve $x_e\left[a_L(y,t_0)\right]$ deduced from Eq.(\ref{xe}) and can be attributed to the spatially-inhomogenous ponderomotive force exerted by the laser field. 

As for the temporal evolution $x_e(t)$ associated to the laser pulse temporal envelop, the prediction of Eq.(\ref{xe}) is shown as a red dashed line in Fig.\ref{sim1}(d), in the case of a fixed ion background: electrons move back to their initial position in the falling edge of the laser pulse, due to their immediate response to the field $a_L(t)$ (Eq.(1)) and the restoring force from the ion background. However, this temporal evolution will be affected when ion motion is taken into account, because $n_0$ then becomes a slow function of time in Eq.(\ref{xe}). The second step of our model aims at determining $n_0(t)$.
   


\subsection*{Ion dynamics}
\par
The charge separation induced by the laser field between the electron and ion populations leads to a quasi-electrostatic field in the plasma, which peaks around $x_e$ and tends to accelerate the ion population located around this position~\cite{Macchi2005}. This acceleration expels the ions from this location, which results in an erosion of the ion density gradient in time. The position $x_i$ of the ion-vacuum boundary thus drifts inward during the laser pulse, and the density $n_0=n(x_i)$ increases in time (Fig.\ref{sim1}(e-f)). 

\par
The so-called hole boring velocity $v_p=dx_i/dt$ of the ion surface can be calculated by writing a momentum flux balance \cite{Denavit1992, Wilks1992, Naumova2009, Schlegel2009,  Robinson2009}. The reflection of the laser beam corresponds to a change in momentum of the field, which is compensated by an opposite change in momentum of the plasma particles. To determine how the light momentum is shared between electrons and ions, we use the same approach as developed independently in \cite{Ping2012}, i.e. we also write the energy flux balance, assuming that the absorbed laser intensity $(1-R)I$ (where $R$ is the plasma reflection coefficient for the laser) is entirely carried away by electrons. The combination of these two balances leads to:
\begin{eqnarray}
x_i(t)&=&2L\ln\left(1+\frac{\Pi_0}{2L\cos\theta}\int_{-\infty}^{t}a_L(t')dt'\right)
\label{xpL}
\end{eqnarray}
with $\Pi_0=(R Z m_e\cos\theta/2 A M_{p})^{1/2}$, where $Z$, $A$ are respectively the average charge state and mass number of the ions, $M_p$ is the proton mass and $m_e$ is the electron mass. The prediction of this equation for $x_i(y,t_0)$ at the laser pulse maximum $t_0$ is shown as a blue line in Fig.\ref{sim1}(b), and fits well the surface of the superimposed ion density map obtained from a PIC simulation with mobile ions. The derivation of  Eq.(\ref{xpL}) shows that this curvature of the ion surface is induced by the spatially-inhomogeneous laser radiation pressure on the PM.

The temporal evolution $x_i(t)$ is represented in Fig.\ref{sim1}(d) by the blue line. As opposed to $x_e$, the ion boundary displacement $x_i$ does not return to its initial value at the end of the pulse. This is because $x_i$ depends on the time integral of $a_L$ (see Eq.(\ref{xpL}), where $a_L$ corresponds to the envelop of the laser field), meaning that it is the cumulated action of the laser field over time that is responsible for the ion dynamics. This results in a progressive change in the ion profile, which in turn affects the electrons dynamics. This coupling is included in our model in a very simple way.

\subsection*{Coupling of electron and ion dynamics}
\par
Due to the erosion of the ion density profile, the laser field now starts pushing the electrons inside the ion background directly from $x_i(t)$, instead of $x_i=0$ initially. Consequently, the position of the electron boundary $x_T(t)$  when ion motion is taken into account is now given by  $x_T(t)=x_i(t)+x_e(t)$ (see Fig.\ref{sim1}(f)).
 In this equation, the value of $x_e(t)$ is also affected by ion motion, because the restoring force induced by the ions initially located between $x=0$ and $x_i(t)$ is suppressed. As explained before, this second effect is accounted for simply by using $n_0=n(x_i(t))$ in Eq.(\ref{xe}).

\par
The temporal evolution of the electron boundary resulting from these coupled dynamics is illustrated in Fig.\ref{sim1}(c). An excellent agreement is obtained between the PIC simulation and the prediction of the full model (black dots). An extensive parametric study of the surface dynamics, using hundreds of PIC simulations, confirms the excellent accuracy ($\leq 5 \%$) of this  model over a broad range of physical conditions (see supplementary information). 

\par
\begin{figure}[b]
\vskip -0.3cm 
\centering \includegraphics [width=8cm]{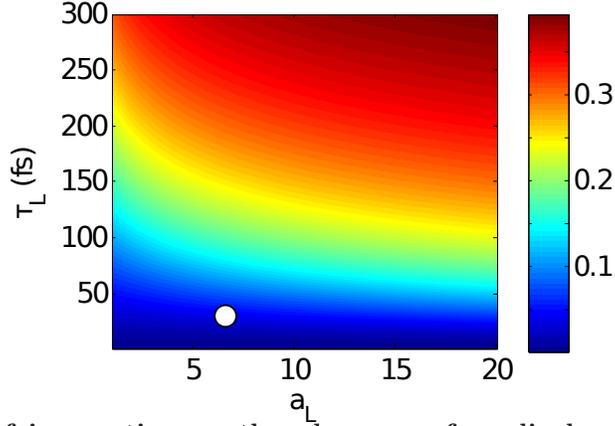}
\vskip -0.5cm 
\caption{\textbf{Influence of ion motion on the plasma surface displacement}. This color map shows the relative change in the plasma electron boundary displacement at the peak $t_0$ of the pulse, $\delta x/x_T=(x_T-x_e)/x_T$, when ion motion is taken into account ($x_T=x_i+x_e(n_0(t))$) and when ions are considered as fixed ($x_e$, Eq.(\ref{xe}) with a constant $n_0$), as predicted by our model. This is plotted as a function of $a_L$ and pulse duration, for a typical value of the density gradient ($L=\lambda/10$). The white dot corresponds to the interaction conditions of the experiment performed with UHI100 (see experimental section).
}
\vskip -0.5cm
\label{sim2}
\end{figure}

Figure \ref{sim1}(d) uses our model to highlight the relative contributions of ion and electron dynamics in the case of the simulation of Fig.\ref{sim1}(c). Despite the brevity of the pulse, the influence of ion motion on the position $x_T$ of the electron boundary becomes significant in the second part of the pulse (beyond $t\approx 10 T_L$). Its main effect is to prevent the electron boundary from moving back to its initial position in the falling edge of the laser pulse, which has observable consequences in experiments, as we will see later. As expected intuitively, the influence of ion dynamics on the total PM surface motion is predicted to become more and more significant as the laser pulse duration increases (Fig. \ref{sim2}).  



\section*{Spatial properties of the reflected beam}

\begin{figure*}[t]
\centering \includegraphics [width=17cm]{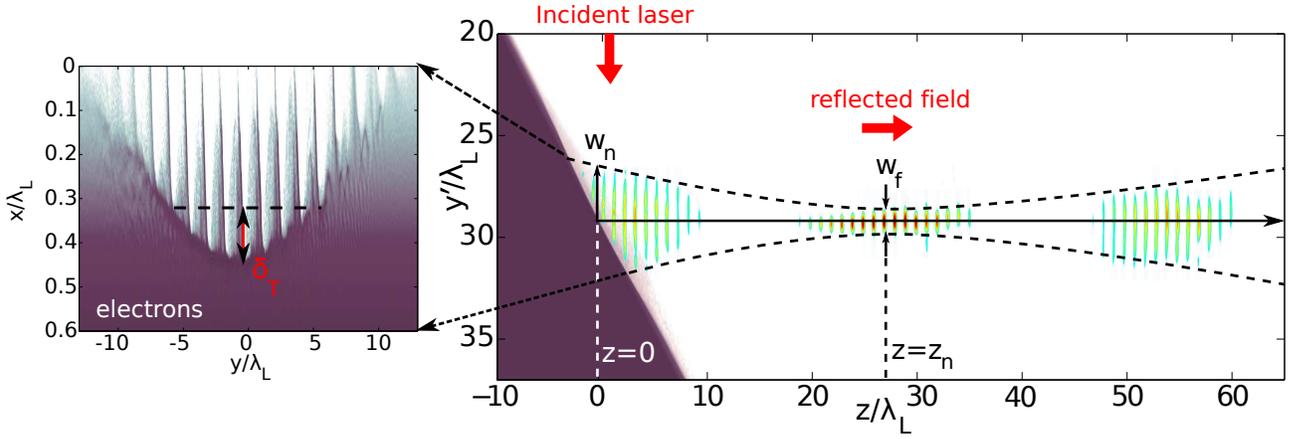}
\caption{\textbf{Focusing of high order harmonics by a curved relativistic plasma mirror in a 2D PIC simulation.} The laser-induced curvature of the PM surface tends to focus the reflected light in front of the PM. Higher harmonic orders are more affected by this curvature, due to their smaller wavelengths. The right panel shows a spatial map of the plasma electron density $n_e$ at the maximum of the laser pulse in dark purple scale. A zoom on the surface is shown in the left panel, which also defines the denting parameter $\delta_T$. The multicolor map shows the intensity $I(y',t)$ of the train of attosecond pulses obtained by filtering ROM harmonics from order 4 to 8, at three different times during its propagation away from the PM. Focusing of this train at a distance $z_n\approx f_p \cos\theta$ from the PM surface is observed, where $f_p$ is the focal length of the curved PM (here $z_n\approx 25 \lambda_L$. i.e. 20 $\mu m$ for $\lambda_L=800 nm$) .
}
\vskip -0.5cm
\label{sim3}
\end{figure*}

The laser-induced denting of the PM leads to a curvature of the wavefronts of the reflected light beam, which tends to focus this beam -including the harmonics generated upon reflection- in front of the surface \cite{dromey_diff,Horlein}. This is clearly observed in Fig.\ref{sim3} on the attosecond pulse train generated by the ROM mechanism, which is focused at a distance $z_n$ from the surface, with a magnification ratio $\gamma_n=w_f/w_n<1$. 
This focusing of the beam naturally tends to increase its divergence. Assuming a Gaussian intensity profile of width $w_n$ for the $n^{th}$ harmonic in the source plane, this divergence is given by (see supplementary information):
 \begin{equation}
\theta_{n}=\theta^{0}_{n}\sqrt{1+\Psi_n^2}
\label{thetan}
\end{equation}
$\Psi_n$ is the PM dimensionless focusing parameter for the $n^{th}$ harmonic, that characterizes the effect of the PM curvature on all spatial properties of the reflected beam: 
\begin{equation}
\Psi_n=\frac{2\pi}{\cos\theta}\left(\frac{w_n}{w_L}\right)^2\frac{\delta_{T}}{\lambda_n}
\label{Psi}
\end{equation}
with $\lambda_n=\lambda_L/n$ the harmonic wavelength. Here $\delta_T$ is defined as $\delta_T=x_T(y=0)-x_T(y=\sqrt{2} w_L)=w_L^2/2 f_p$ (Fig.\ref{sim3}(left)), i.e. it is the difference between the surface position at the center of the focal spot $y=0$, and its position at $y=\sqrt{2}w_L$ (with $w_L$ the half spatial width at $1/e$ of the laser field amplitude). In Eq.\ref{thetan}, $\theta^0_n=\lambda_n/\pi w_n$ is the divergence that would be obtained in the absence of surface curvature, i.e. imposed by diffraction from the source plane. It can be expressed as a function of laser divergence $\theta^0_n=\theta_Lw_L/w_nn$.

In Eq.(\ref{thetan}), each term of $1+\Psi_n^2$ corresponds to a different physical limit. If $\Psi_n \ll 1$ (e.g. $\delta_T \ll \lambda_n$ or $w_n \ll w_L$), surface curvature has a negligible effect on the spatial properties, which are determined only by the beam diffraction from the source plane. On the opposite, if $\Psi_n \gg 1$, the focusing induced by the PM imposes the beam divergence, leading to $\theta_n \rightarrow \Psi_n \theta^0_n \gg \theta^0_n$.
$\Psi_n$ is in principle a function of time. However, our model shows that after a fast transient of less than five laser periods only, $\delta_{T}$ and hence $\Psi_n$ weakly vary in time (black dots in Fig.\ref{sim1}(d)). As a first approximation, we therefore neglect its temporal variation in our study of the spatial properties of the reflected beam.
 
This model for the reflected beam properties has been successfully compared with a series of 2D PIC simulations (see supplementary information). 
In the interaction conditions corresponding to the present state of the art of femtosecond lasers ($a_L \lesssim 10$, $L \lesssim \lambda_L/5$), it predicts $\delta_T \approx 0.1 \lambda_L$ (80 nm for $\lambda_L=800$ nm) and $\Psi_n\approx 0.6n$ typically. The effect of surface curvature thus already becomes significant for harmonic orders $n \gtrsim 3$. We now turn to an experimental investigation of the spatial properties of such harmonics, to validate the model, and show what insight it provides on HHG, and more generally on the physics of plasma mirrors.

\section*{Experimental study}



The experiment was performed on the UHI100 laser of IRAMIS (CEA, France), that delivers 25 fs pulses with a peak power of up to 100 TW and an ultrahigh temporal contrast (see Methods section). This beam was focused in $p$-polarization to a spot size of 4 $\mu m$ on a silica target, reaching an estimated peak intensity of $6.7\times10^{19}$ $W/cm^2$ ($a_L=5.6$), thus producing a relativistic plasma mirror. 
The density gradient scale length $L$ at the PM surface was varied by using a small controlled prepulse, intense enough to create a plasma ($I=10^{16}$ $W/cm^2$) at an adjustable delay $\tau$ ($0\leq\tau\leq 2$ $ps$) before the main pulse. The value of $L$ was determined experimentally using time-resolved interferometry \cite{geindre_fdi, KahalyGradient}.

\subsection*{Measured spatial properties of harmonic beams}

Under these conditions, high-order harmonics are produced in the reflected beam by the ROM mechanism \cite{article_NP,dromey_NP}, and two diagnostics were used to characterize the spatial properties of the resulting harmonic beam in the far field (see Fig.\ref{exp1}(a-b) and Methods section). 
   The spectrally-resolved divergence, extracted from images such as shown in Fig.\ref{exp1}(a), is presented in Fig.\ref{exp1}(c-d)  as a function of harmonic order and of the density gradient $L$ at the PM surface. The full lines show the results of the model. The only two unknowns of the model are the plasma reflectivity $R$ (used only to calculate the ionic contribution to the surface curvature), and the ratio of harmonic and laser source size $w_n/w_L$ (used to deduce the harmonic divergence from the surface curvature). These are however not used as free parameters to fit the data, but are directly extracted from 2D PIC simulations performed in the physical conditions of the experiment (see Methods section). This provides $R\approx0.7$ and $w_n/w_L \approx 0.5 $ for the $25^{th}$ harmonic. Parametric studies (see supplementary information) show that these values hardly change over a broad range of interaction conditions ($a_L$ and $L$). In addition, we note that $R$ hardly influences the results, since it only affects ion motion and appears in a square-root in Eq.2. These curves are in remarkable agreement with the measurements, thus validating the model and showing it can be used to gain insight on the physics involved in this experiment. Note that this agreement was obtained without introducing any additional `intrinsic phase' $\varphi$, such as the one described by An der Br\"{u}gge et al \cite{anDerBrugge}. Our model actually suggests that this phase is simply given by $\varphi = 2\pi x_{e}/\lambda_{L}$ where $x_{e}$ is the electron denting provided by Eq.(\ref{xe}), and is thus implicitly included in our analysis. This expression exactly predicts the scaling of $\varphi$ obtained in \cite{anDerBrugge} for normal incidence and a step-like plasma surface in the limit of ultra-relativistic intensities.

Comparing the measured divergences with those that would be obtained by diffraction from a flat PM for the same source size [black dashed lines in Fig.\ref{exp1}(c-d)] shows that the harmonic divergence is close to this limit when $L$ is small, but is then very significantly increased by the PM curvature, here by a factor of up to 3, for the typical gradients that optimize the ROM conversion efficiency ($L\approx 0.05$ to $0.1\lambda_L$) \cite{FabienJPB2010,Rodel_grad,KahalyGradient}. This analysis provides a clear indication of the focusing of the harmonics in front of the PM, due to its surface curvature. The measurements of Fig.\ref{exp1}(d) show that this focusing increases with the gradient scale length $L$, as expected from the model, since a longer gradient leads to a softer restoring force from the ion background, and hence to a larger surface denting $\delta_T$. 

\par
The laser pulse duration used in this experiment is so short that ion motion has little influence on the PM curvature, and hence on the harmonic divergence (white dot on Fig.\ref{sim2}). 
According to Fig.\ref{sim1}(d), it however significantly changes the temporal dynamics of the surface in the falling edge of the pulse. We now demonstrate that this can lead to observable effects in the experiment, by considering the spectral properties of the harmonics. 

\subsection*{Temporal dynamics and Doppler effect}

After a fast initial transient where the denting $\delta_T$ strongly varies, the temporal evolution of the field envelop $a_L(t)$ only leads to a weak residual drift of PM surface during the pulse (black dots in Fig.\ref{sim1}(d)), with typical velocities of the order of $0.01 c$ according to our model. This motion appears as a slow drift on the femtosecond time scale, that combines with the fast relativistic oscillation of the plasma surface at the laser frequency responsible for HHG (see Fig.\ref{sim1}(c)). This results in a Doppler shift on the reflected light, which scales linearly with harmonic order $n$, and thus gets measurable for large enough values of $n$.

Since the ion dynamics affects the temporal evolution of the PM surface, it can potentially influence this Doppler effect. This is confirmed by a comparison of PIC simulations performed with fixed (Fig.\ref{sim4}(a)) and moving ions (Fig.\ref{sim4}(b)). In the case of fixed ions, the plasma surface moves inward in the rising part of the pulse, leading to a Doppler redshift, and then moves outward in the falling part, leading to a Doppler blueshift. If strong enough, this effect leads to harmonics with a double peak structure \cite{Behmke_at}, clearly observed in Fig.\ref{sim4}(a). In contrast, when ion motion is allowed in the simulation, the irreversible erosion of the ion density gradient prevents the electron boundary from moving back to its initial position when the laser intensity decreases. This naturally suppresses the Doppler blueshift, and only a Doppler redshift is observed, in the interaction conditions considered here.

Turning back to the experiment, Fig.\ref{exp2}(a) shows a zoom on the spatio-spectral distribution of the $23^{rd}$ harmonic, measured in a typical shot. It is very similar to the PIC results of Fig.\ref{sim4}(b), and only a red shift is observed: according to the previous discussion,  this is a signature of ion motion. Fig.\ref{exp2}(b) shows that the Doppler shift at the center of the beam increases with the density gradient scale length $L$, which is consistent with the stronger curvature of the PM for larger $L$. This dependence is quantitatively reproduced by our model, when both ion and electron dynamics are taken into account using the same parameters as in Fig.\ref{exp1}. Thus, although ion dynamics does not affect the spatial properties of harmonics in our experimental conditions, it has a clear signature in the spectral domain, which validates the ionic part of our model.

\section*{Discussion and outlook}
\begin{figure*}[l]
\vskip -0.2cm 
\centering \includegraphics [scale=0.85]{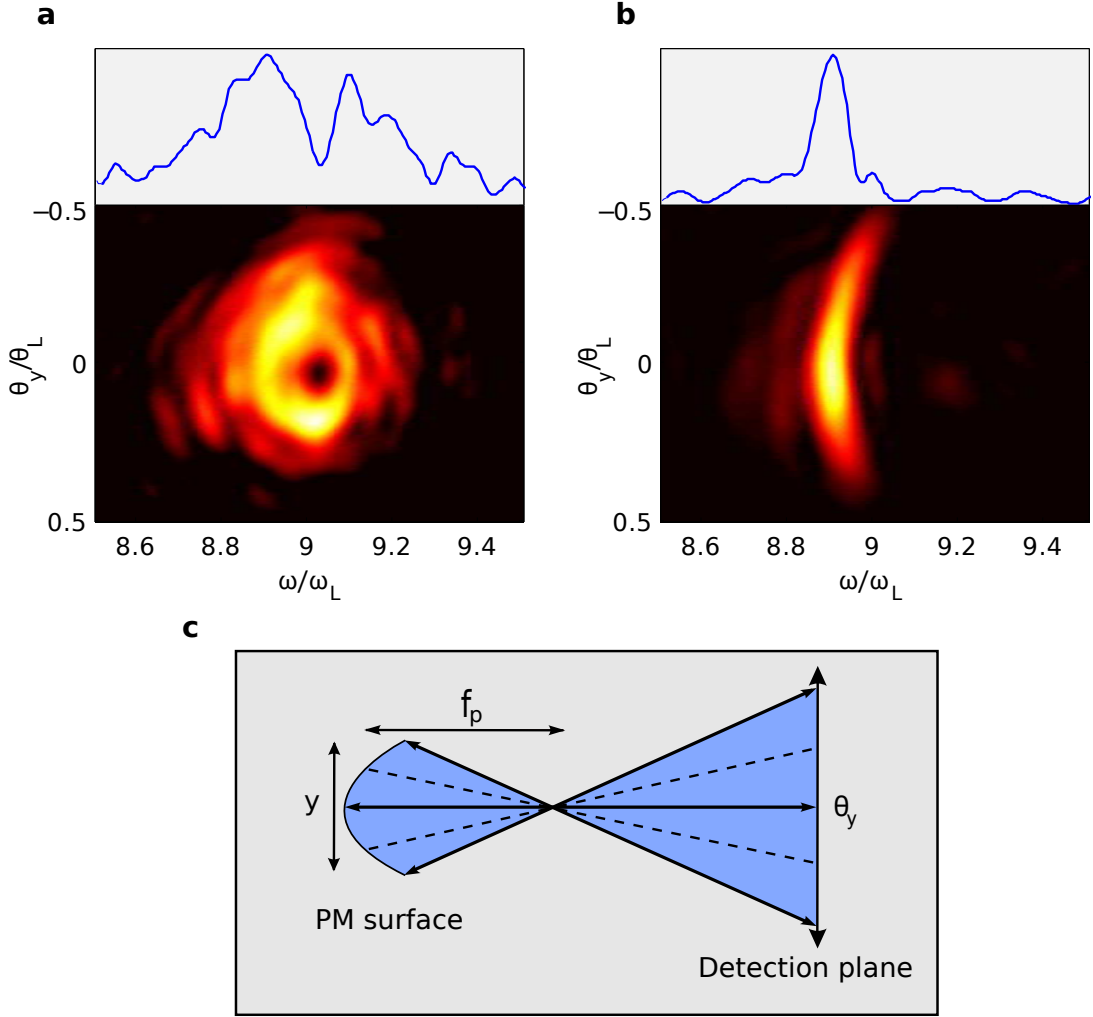}
\caption{\textbf{Doppler shift of high order harmonics.} The two images show the angularly-resolved spectra of the $9^{th}$ harmonic, obtained from 2D PIC simulations for $a_L=8$ and $L=\lambda/8$, considering either fixed (a) or mobile (b) ions. The curves in the upper panels shows line outs of the spectra at the center of these beams ($\theta_y=0$). In these interaction conditions, the PM curvature is strong enough to lead to a geometrical mapping of the PM surface onto the propagation angle (sketch in panel (c)). The center of the focal spot, where the surface recession velocity is the largest and the Doppler redshift the strongest, is mapped to the center of the far-field beam. This explains the shape of the distributions in (a) and (b), where the Doppler shifts are always larger at the center of the beams. 
}
\vskip -0.55cm
\label{sim4}
\end{figure*}

We have presented a simple analytical model for the spatial properties of light beams reflected by relativistic plasma mirrors, in excellent agreement with both PIC simulations and experimental results. It provides insight into the respective roles of ion and electron dynamics, and into the spatial and spectral properties of harmonics generated in the reflected beam. Combined with this model, these harmonics now constitute a direct and powerful diagnostic of the femtosecond motion of the PM surface, with spatial resolution within the laser focal spot (Fig.\ref{sim4}(c)). Measurements schemes such as photonic streaking \cite{KTK} will potentially also provide temporal resolution within the laser pulse envelop.

\begin{figure*}[r]
\vskip -0.2cm 
\centering \includegraphics [width=10cm]{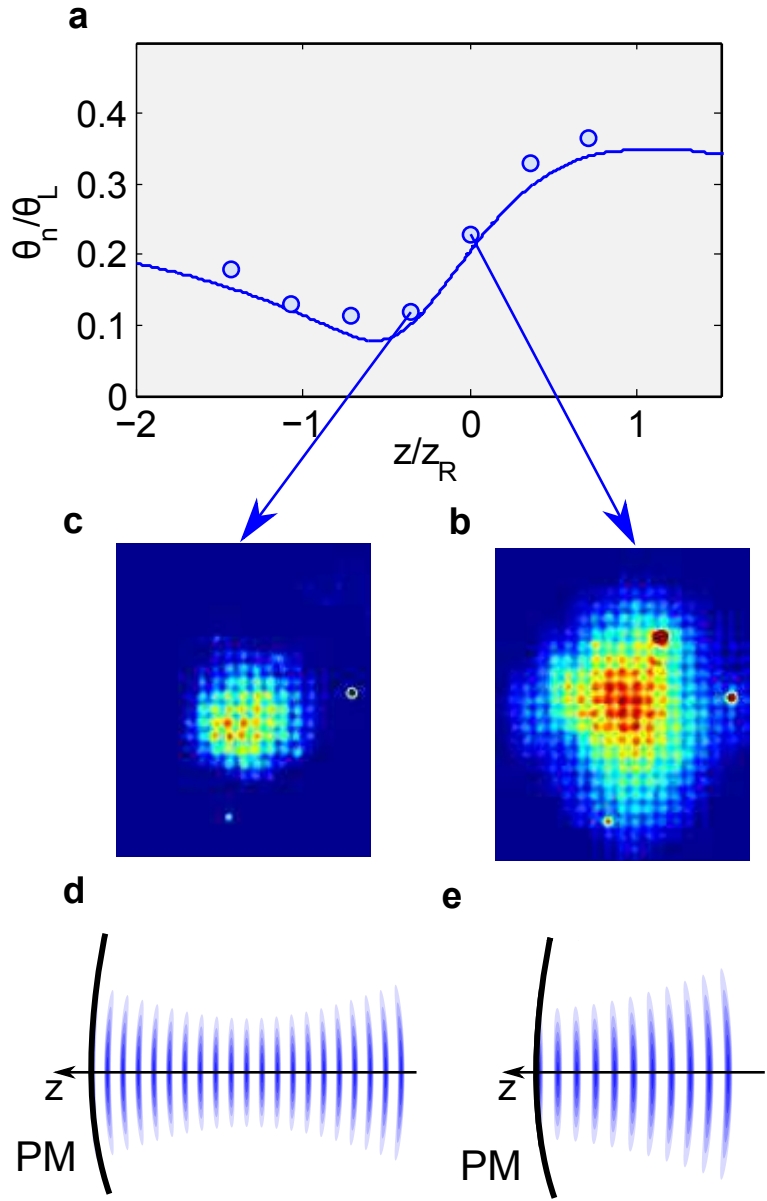}
\caption{\textbf{Control of the harmonic divergence from a relativistic plasma mirror}. The blue dots in panel (a) show the divergence of the $25^{th}$ harmonic measured in our experiment, as a function of the distance $z$ between the laser best focus and the target surface. The full line shows the prediction of the model. The two images in (b) and (c) show the 2D spatial profiles of the harmonic beam measured at the two focusing positions indicated by the arrows, sketched in (d) and (e). A slightly curved, diverging laser wavefront ($z<0$) (panel (d)) compensates the effect of the laser-induced PM curvature, thus reducing the harmonic beam divergence to a value that is imposed by diffraction from the source plane. All experimental data points correspond to a single laser shot. 
}
\vskip -0.5cm
\label{exp3}
\end{figure*}

This model will be instrumental in designing future applications of plasma mirrors, in particular for attosecond science. It can for instance be used to determine what laser pulse duration is required to generate isolated ROM attosecond pulses using the lighthouse effect \cite{fabien_light,pharexp}. In this perspective, as well as in most applications where the reflected beam is manipulated or used in the far-field, being able to control and minimize the attosecond beam divergence is essential~\cite{Horlein}, which requires mitigating the effect of the laser-induced PM curvature. Figure \ref{exp3} provides the first experimental demonstration in the relativistic regime of a very simple scheme for such a control~\cite{PRL_phase}: by using a driving laser-beam with a slightly diverging wavefront on target, the effect of the PM curvature on the attosecond beam can be compensated, leading to a divergence close to the one that would be obtained for a flat mirror, reduced by a factor of more than 2 compared to the one obtained at best focus. 

In other applications, PM will prove useful to focus the reflected beam, and boost the peak intensity of the fundamental laser frequency~\cite{Motoaki} or its harmonics~\cite{Gordienko_focusing, Gonoskov2011}. This can be achieved using either curved substrates, or the natural light-induced PM curvature described in this work, which typically leads to magnification factors $\gamma_n=w_f/w_n\approx 0.1$ for $n \geq 10$ in the interaction regime considered here. In either case, the understanding of the laser-induced PM surface dynamics provided by this work will be essential.





\par
The research leading to these results has received funding from the European Research Council (ERC Grant Agreement No. 240013) and Laserlab-ALADIN (Grant No. 228334). This work was performed by using HPC resources from GENCI-CCRT/CINES (Grant No. 2012-056057).\\

\begin{figure*}[t]
\centering \includegraphics [scale=0.85]{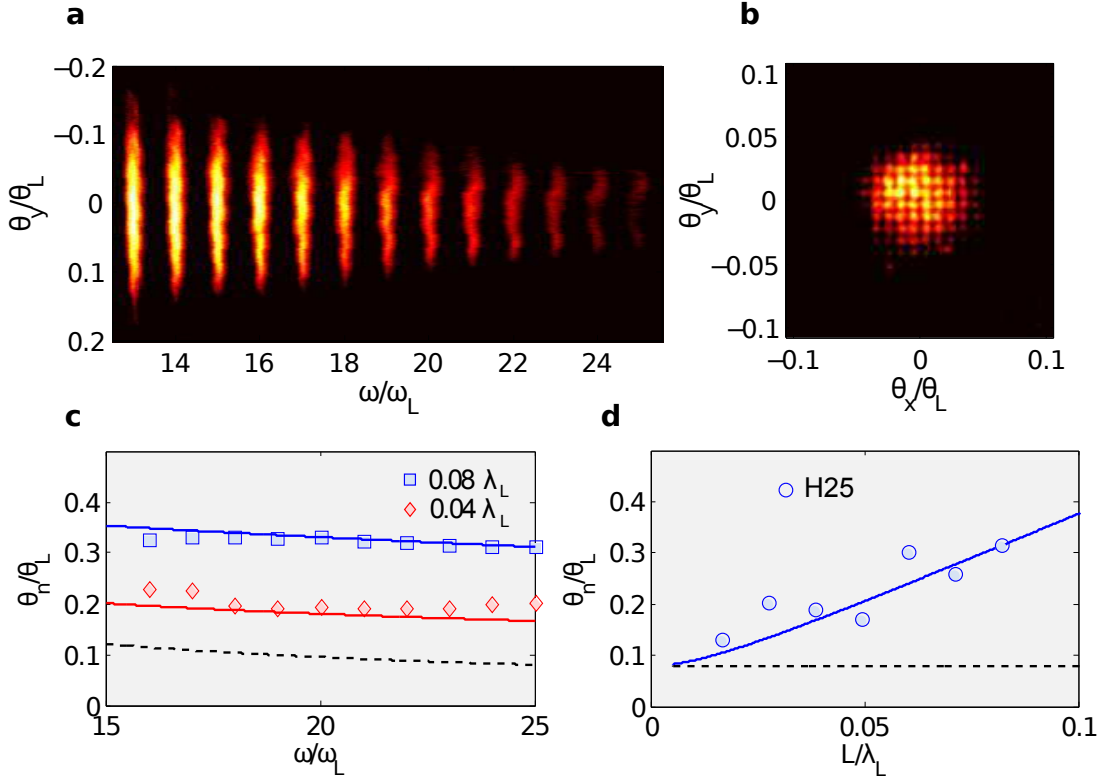}
\vskip -0.3cm 
\caption{\textbf{Measurements of high-order harmonics beams produced by a relativistic oscillating mirror.} (a-b) Typical raw images obtained with the UHI100 laser. See methods section for a description of the diagnostic instruments. Image (a) shows the angularly-resolved harmonic spectrum measured in the far-field, for a peak intensity of $I=3.5\times10^{19}$ $W/cm^2$ ($a_L=5.6$), and an initial density gradient $L=\lambda_L/20$. The apparent decrease of the harmonic divergence with order is mainly an effect of the 2D color map. Image (b) displays the full far-field spatial profile of the beam corresponding to the superposition of harmonics 20 to $\sim35$, measured in similar interaction conditions. The shadow of the supporting mesh of the thin $Si$ filter used to select a group of harmonics is clearly observed. 
From these images, quantitative information on the harmonics spatial properties can be extracted. Panel (c) thus shows the spectrally-resolved divergence (in units of laser divergence $\theta_L$, with $\theta_L=200$ mrad in our experiment) as a function of harmonic order, for two values of the density gradient $L$. The plot in (d) is the divergence of the $25^{th}$ harmonic as a function of $L$. In both panels, the full lines show the results of the model. The divergence $\theta_n^0=(w_L/w_n)(\theta_L/n)$ that would be imposed by diffraction from the same source size in the absence of the laser-induced PM curvature, is shown as the dashed lines in (c) and (d). A fully-consistent set of parameters was used for all curves. All experimental data points correspond to a single laser shot.  
The two shots displayed in panel (c) correspond to the data points for which the absolute value of the divergence is in best agreement with the model in panel (d).
}

\vskip -0.5cm
\label{exp1}
\end{figure*}

\begin{figure*}[r]
\vskip -0.2cm 
\centering \includegraphics [scale=0.85]{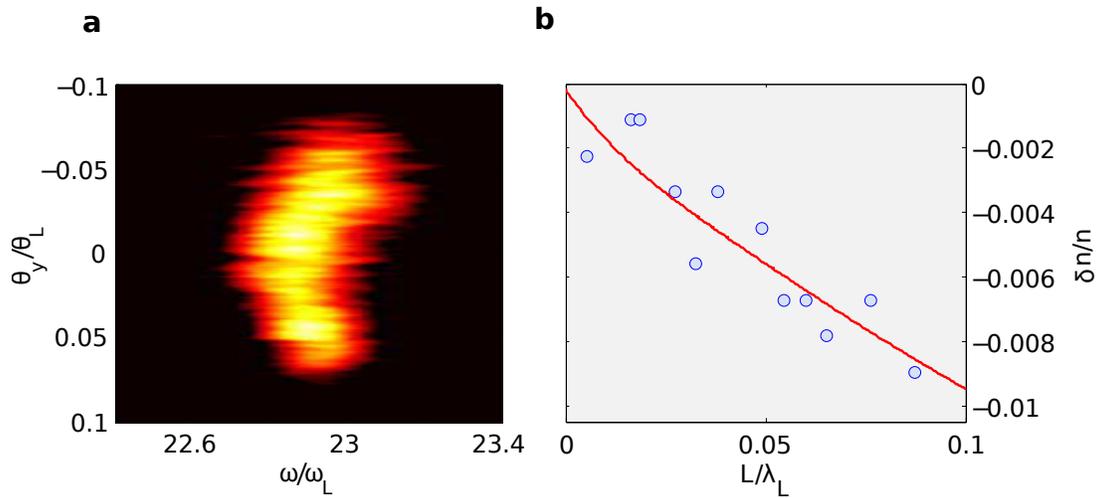}
\caption{\textbf{Measured Doppler shift of an individual harmonic}. The image in (a) shows the measured angularly-resolved spectrum of the $23^{rd}$ harmonic, for $a_L=5.6$ and $L=0.044 \lambda_L$. It is a zoom on a measured image such as displayed in Fig.\ref{exp1}(a). An angle-dependent Doppler redshift is observed, like in the PIC simulation with moving ions of Fig.\ref{sim4}(b), due to the position-to-angle mapping resulting from the PM curvature. Panel (b) shows the measured Doppler shift at the center of the beam, as a function of the density gradient scale length $L$. The full line shows the results of the model. All experimental data points correspond to a single laser shot.
}
\vskip -0.5cm
\label{exp2}
\end{figure*}

\textbf{Methods}\\
\textbf{Simulations}\\
\footnotesize {We used the PIC codes EUTERPE in 1D, and CALDER in 2D, to confront our model to simulations. In all simulations, we considered a $p$-polarized laser pulse of amplitude $a_L$ impinging with an angle $\theta$ on a plasma density profile that has a maximum density of $200n_c$, and an initial exponential density gradient of scale length $L$. The laser field is injected in the simulation box through boundary conditions. In 1D, we account for the oblique incidence by performing all the calculations in the boosted frame. The size of the simulation box is $30\lambda_L$, with a mesh size of $6.7\times10^{-4}\lambda_L$, the time step is $4\times10^{-4}T_L$, and we used 500 particles/cell. A typical calculation requires 24 hours on 1 CPU. In 2D, the simulations parameters are: a simulation box of $30\lambda_L\times 40\lambda_L$, with a mesh size of $2.8\times10^{-3}\lambda_L$, a time step $2\times10^{-3}T_L$ and 20 macroparticles/cell. A typical calculation requires 24 hours on 512 CPUs. All simulation results presented in the main text are from 2D simulations with CALDER. The results of 1D simulations with EUTERPE are presented in the online supplementary material.}\\
\textbf{Experiment}\\
\footnotesize{The experiments are performed using the UHI100 Ti:sapphire laser, that delivers 25 fs FWHM pulses centered at 800nm. The $ps$ pulse contrast is improved to more than $10^{12}$ using an antireflection coated double plasma mirror set up. The high contrast p-polarized laser beam is then aberration corrected using an adaptive optical system, and focused on an optically flat target at an incidence angle of $55^{o}$ using an off axis parabola. A small fraction of the main beam is picked up for the prepulse and is focused to generate a preplasma. The prepulse focal spot is 5 times larger than that of the main beam allowing homogeneous density gradient all across the HHG source. The controlled delay between the prepulse and the pump beam determines the initial gradient scale length $L$ which is measured using time-resolved interferometry. The harmonic beam produced by the main laser pulse on the gradient-controlled plasma mirror is spectrally dispersed and angularly resolved using a 1200 lines/mm varied line spacing XUV grating (Shimadzu 30-002), and is detected on a 69x88mm rectangular micro channel plate (MCP) (Fig.\ref{exp1}(a)). For the 2D spatial diagnostics (Fig.\ref{exp1}(b)), the reflected beam is spectrally filtered by the combination of two silica plates used at grazing incidence and AR-coated at the laser wavelength, and a high pass 250 nm thick Si filter, and then detected using another MCP. The MCPs are coupled to phosphor screens imaged on 12 bit CCD cameras.}\\
\textbf{Fit of experimental data with the model}\\
\footnotesize{We extracted the harmonic source size $w_n$ from the results of 2D PIC simulations performed in the physical conditions of the experiment, and obtained $w_n/w_L=0.72-9.10^{-3}.n$ for orders $n$ between 5 and 25. For the theoretical curves shown in Fig.\ref{exp1}, $w_n/w_L$ thus varies from $0.59$ for $n=15$ to $=0.5$ for $n=25$. However, the curves are hardly changed if a constant source size of $w_n/w_L=0.5$ is used for this entire spectral range. The same values of the source size ratio were used for all gradient scale lengths $L$, as suggested by PIC simulations.}


\pagebreak

\end{document}